# COVID-19 vs Social Media Apps: Does Privacy Really Matter?

Omar Haggag*, Sherif Haggag†, John Grundy*, Mohamed Abdelrazek†
*Faculty of Information Technology, Monash University, Australia
{omar.haggag, john.grundy}@monash.edu
†School of Information Technology, Deakin University, Australia
{sherif.haggag, mohamed.abdelrazek}@deakin.edu.au

*Abstract*—Many people around the world are worried about using or even downloading COVID-19 contact tracing mobile apps. The main reported concerns are centered around privacy and ethical issues. At the same time, people are voluntarily using Social Media apps at a significantly higher rate during the pandemic without similar privacy concerns compared with COVID-19 apps. To better understand these seemingly anomalous behaviours, we analysed the privacy policies, terms & conditions and data use agreements of the most commonly used COVID-19, Social Media & Productivity apps. We also developed a tool to extract and analyse nearly 2 million user reviews for these apps. Our results show that Social Media & Productivity apps actually have substantially higher privacy and ethical issues compared with the majority of COVID-19 apps. Surprisingly, lots of people indicated in their user reviews that they feel more secure as their privacy are better handled in COVID-19 apps than in Social Media apps. On the other hand, most of the COVID-19 apps are less accessible and stable compared to most Social Media apps, which negatively impacted their store ratings and led users to uninstall COVID-19 apps more frequently. Our findings suggest that in order to effectively fight this pandemic, health officials and technologists will need to better raise awareness among people about COVID-19 app behaviour and trustworthiness. This will allow people to better understand COVID-19 apps and encourage them to download and use these apps. Moreover, COVID-19 apps need many accessibility enhancements to allow a wider range of users from different societies and cultures to access to these apps.

## I. INTRODUCTION

Billions of people are using a wide range of Social Media mobile apps every day. As shown in [1], there is a massive spike in using Social Media platforms during the various lockdowns associated with the COVID-19 pandemic. For example, more than 2.7 billion people used Facebook every month in the second quarter of 2020 [2], and WeChat mobile app is actively used by over 1.2 billion people every month [3]. Another relevantly new application called TikTok was launched to the international market in September 2017. TikTok has around 800 million active users worldwide nowadays [4]. Social media apps have played a great role in entertaining and reconnecting people, friends, and families during the lockdown, isolation, and social distancing. Productivity apps have also seen a massive jump in usage [5]. For example, Grammarly, which helps people to fix their grammar, spelling, and punctuation, has over 10 million downloads [6].

Since the beginning of 2020, the COVID-19 pandemic has completely changed the way we live and work across the globe [7]. Information and Communications Technology (ICT), Teleworking Technology, and digital devices have played a huge role in this major change. Moreover, these technologies have greatly impacted the economy by saving jobs of people and allowing businesses in different sectors to continue operating during lock-downs. It has forced people to communicate and deal with each other via online applications such as ZOOM, Skype, and Microsoft Teams [8].

A new type of mobile app has appeared, required or advised by governments and health officials [9]. These *COVID-19 applications* are an attempt by governments and technology companies to minimise the spread of the virus. Besides traditional, manual contact tracing techniques, COVID-19 apps assist authorities to rapidly find and track individuals that have been exposed to the virus [10], [11]. They also help to raise awareness about the virus via educational information, including symptoms and how it can be spread. The success of contact tracing apps depends on the percentage of the public using these apps regularly [12]. The higher rate of adoption, the higher percentage these tracing apps can help [13]. Contact tracing apps collect data by tracing the movements of people, identifying where they have been, with whom they have been in close contact, and where they may have propagated the virus. These data are usually encrypted, anonymous, and stored historically rather than in real-time just to protect the privacy of the users [14]. Without contact tracing apps, finding these nearby contacts depends on individuals having the ability to remember everybody they have been in close contact with.

However, many people are worried about using COVID-19 apps [15]. There are several reasons behind this, including privacy issues, lack of trust and ethical concerns [15]. Also, several media platforms play a huge role in swinging reality to scare and panic the public and steer them away from using COVID-19 apps [16]. On the other hand, Social Media & Productivity mobile apps have lots of well-demonstrated and publicised privacy and ethical issues. However, the public are still using them, often without any apparent concerns, and even at a higher rate during the pandemic [1], [17].

To better understand this behaviour, we carried out a detailed, manual analysis of the privacy policies, terms & conditions, and data use agreements in COVID-19, Social

Media & Productivity apps. We also carried out an automated analysis of user reviews of COVID-19, Social Media & Productivity apps from both Google Play and App Store. Review analysis covered five aspects – privacy, stability, advertising, user requests, and uninstallation.

**The key contributions of this work are:**

- A detailed analysis of privacy-related policies of common COVID-19 and social media apps, and we find the later are far more prone to privacy and data mis-use than COVID-19 apps.
- An analysis of nearly 2 million user reviews of common social media apps and COVID-19 contact tracing and education apps, and we find the later suffer from far more severe accessibility and stability issues.
- Evidence-based recommendations for developers and promoters of COVID-19 apps to ensure they are technically improved but also better understood by users.

## II. MOTIVATION

As the COVID-19 pandemic continues, the number of daily cases has been massively expanding [18]. As we write this late 2020, Europe is facing the second wave of the virus and no official or final vaccine has yet been produced [19]. The authors of a key recent study [13] indicated that COVID-19 apps can play a very important role in stopping the global pandemic if 60% of the public download and start using COVID-19 mobile tracing apps. Moreover, another study indicated that even lower percentages of using the apps will help in fighting the virus and decrease its propagation [12]. Many health officials and experts are advising the public to download and start using COVID-19 mobile apps [20]. This helps them to fight the virus in several ways since COVID-19 tracing mobile apps are more successful and effective if many people start using them [12], [13].

> **Facebook**: "We use location-related information such as your current location, where you live, the places you like to go, and the businesses and people you're near. Location-related information can be based on things like precise device location, IP addresses, and information from your and others' use of Facebook Products (such as check-ins or events you attend) [21]"
> **TikTok**: "We automatically collect certain information from you when you use the Platform, including internet or other network activity information such as your IP address, geolocation-related data, unique device identifiers, browsing and search history (including content you have viewed in the Platform) [22]"
> **WeChat**: "Location Data is information that is derived from your GPS, WiFi, compass, accelerometer, IP address, or public posts that contain location information. Location information will be disclosed (either to us, to other users, or both) [23]"
> **Grammarly**: "As a rule, Grammarly employees do not monitor or view your User Content stored in or transferred through our Site, Software, and/or Services, but it may be viewed if we believe the Terms of Service have been violated and confirmation is required [24]"

Fig. 1: Excerpts of privacy policies of common social media apps

However, it is obvious that many people are worried about downloading COVID-19 apps. This seems to be because they believe it will help governments and tech companies to better track them and they will become more exposed. Privacy issues in COVID-19 apps are usually raised by people (users), Media, and Human Rights organisations who are usually against the governments who developed these apps [16], [25], [26].

Different governments around the world are enforcing social distancing and lockdown measurements as advised by health officials and experts [12], [13]. These lockdown measurements are leading people to become more active online. Many people are using Social media apps more frequently to help them re-connect with their family, friends, and others [1]. For Social Media apps, privacy issues are usually raised by governments or in some cases the app developers themselves, such as YouTube and TikTok. Most users of these apps are not highly involved in those discussions [27], [28]. There are huge privacy and ethical concerns for Social Media & Productivity apps which can be dangerous to individuals and even governments. For example, both Australian and US governments are trying their best to prohibit Chinese-owned Social Media apps such as TikTok and WeChat. This was because they are claimed to collect large amounts of personal data about users, which helps them build a strong understanding of those users [29]. These collected data can be used in an unethical way, such as the recent example in the US election where Facebook and YouTube were used to potentially swing the US election, which benefits opposing foreign governments [30]. On the 19 of September 2020, the US government decided to ban any new downloads for TikTok and WeChat to protect the privacy of their citizens from being exposed [31].

By looking at the privacy policies and data usage agreements for the most used Social Media and Productivity apps, we can see many privacy issues that many users are not very aware of them. Figure 1 shows parts from the latest privacy policies showing that users' data comes first not their privacy. In contrast, an analysis of the privacy policies and data usage agreements for most COVID-19 mobile apps shows that users' privacy comes first not their data. Figure 2 shows a few parts of their latest privacy policies. This motivated us to start investigating and understanding users' reviews.

> **COVIDSafe**: "No location data (data that could be used to track your movements) will be collected at any time. No user will be able to see the contact data stored on their device as it will be encrypted. Contact data stored on a device will be automatically deleted after 21 days [32]"
> **Corona-Warn-App**: "The system does not know who this person met, nor when or where the encounter took place. No identification is required for the app to identify that an encounter took place. Furthermore, the open-source approach of the source code enables the community to review the functionality of the app and to leave comments [33]"

Fig. 2: Some excerpts of privacy policies of some common COVID-19 apps

No work has yet been done to understand what motivates people to use social media apps and yet refuse to use COVID-19 apps. In this work, we aim to understand this contradiction between what the privacy policies stated and end users' perception. Understanding this behaviour is very important for public health officials, experts, and technologists as it will help them to improve app design, techniques, and privacy policies. Subsequently, they will release apps updates that address the public concerns based on these user reviews. This will increase the download and adoption rates of COVID-19 mobile tracing apps. In our study, we review and analyse how personal data is handled in COVID-19 and Social Media mobile apps based on their privacy policies and their terms & conditions. We then try to understand people's behaviour by analysing user reviews for both COVID-19 and Social Media mobile apps. This user reviews analysis includes privacy, stability, advertising, and uninstallation related issues raised by users in both Google Play and App Store apps.

## III. METHOD

We carried out two types of analysis to compare COVID-19 apps and Social Media & Productivity apps. First, we did a manual analysis of their privacy policies, terms & conditions and data use agreements. Second, we did a large scale automated analysis of nearly 2 million user reviews for COVID-19, Social Media & Productivity apps.

### A. Dataset

For the privacy policies analysis, we manually analysed 42 official COVID-19 mobile apps from different countries and compared them to Facebook, TikTok, WeChat and Grammarly mobile apps. We chose Facebook, TikTok, WeChat, and Grammarly to represent commonly used Social Media & Productivity mobile apps, as they are the most common apps with the highest number of active users and downloads and each app has a different usage purpose. All COVID-19 apps included in the study are named in Table III.

For the user review analysis and to ensure diversity, we ranked official COVID-19 contact tracing mobile apps by both downloads and adoption rates. We then selected the top used apps across different continents around the world. This resulted in choosing 21 COVID-19 mobile apps to carry out user review analysis as shown in Table I. For Social Media & Productivity mobile apps, we decided to do our user review analysis on Facebook, TikTok, WeChat, and Grammarly. We also included an app "COVID Symptom Study", designed by doctors, scientists and technologists in the United Kingdom. They try to address privacy concerns raised by the public against traditional COVID-19 contact tracing apps, as the app only requests users to answer some questions about the virus and shares important information about the virus to users.

### B. RQ1 – How is personal data handled by COVID-19 and Social Media mobile apps?

To address the privacy and data use agreement issues, we developed a set of questions about app privacy and data management policies. This includes what kind of user data are produced, stored, exchanged, and transmitted to third parties and between whom these data are shared and why. These questions will allow us to understand how users' data are handled by COVID-19 & Social Media mobile apps.

| App Name | Country | | |
|---|---|---|---|
| Aarogya Setu | India | Immuni | Italy |
| BeAware | Bahrain | NHSCOVID | UK |
| COCOA | Japan | NZ COVID Tracer | New Zealand |
| Corona-Warn-App | Germany | Rakning C-19 | Iceland |
| COVIDSAFE | Australia | Seha Masr | Egypt |
| COVID-19 Gov | Pakistan | Smittestopp | Denmark |
| Ehteraz | Qatar | StopCovid | France |
| eRouska | Czech | Stopp Corona | Austria |
| HaMagen | Israel | SwissCovid | Switzerland |
| Hayat Eve Sığar | Turkey | Tabaud | Saudi Arabia |
| | | Trace Together | Singapore |

TABLE I: COVID-19 apps used in our user reviews analysis

*1) Is it an Open Source application?:* Open source apps have all application code publicly available for anyone to download and review. This gives users the opportunity to understand precisely how the app is working, what algorithms does the app use and how does the app handle user data.

*2) Are you forced to download or use it?:* We wanted to know if downloading the app is mandatory or it is voluntarily downloaded by users. Some governments are forcing their citizens and residents to download some apps for specific reasons.

*3) Is your data used out of the app scope?:* This is a critical point to understand ethical and privacy issues of an app. If the app is using users' data for other reasons that it is supposed to, then there are serious issues.

*4) Is your data deleted when it's no longer needed?:* From a privacy and security perspective, mobile apps should delete all the user data once it is no longer needed. The data collected by any app should not last forever.

*5) Does the app collect more data than it actually needed?:* Some mobile apps collect more data than they actually need, not just the data that the user enters. Some of these data are collected based on user behaviour and activity.

*6) Is your identity masked?:* Some mobile apps mask user's data in order to protect their identity. However, some applications do not and anyone accessing the data can discover user identity and behaviour with the app.

*7) Can the App access everything you type?:* Some mobile apps have access to record and monitor every single letter users type, which is a major ethical and security issue.

### C. RQ2 – What are the key issues raised by the users of COVID-19 and Social Media & Productivity apps as evidenced in their users' reviews?

To answer this question, we carried out a large scale user review analysis to investigate how privacy, stability, and advertising issues impacted users' ratings. We analysed users' requests and inquiries and how they affected the submitted ratings. We also reviewed the main reasons why users decide to uninstall or delete these apps.

**Our app review analysis and classifying tool:**
We developed a tool that downloads users' reviews, translates them to English, classifies them into five main aspects, and summarises the results and findings. The five aspects used in our classification are explained in Table II. We used our tool to download and analyse just under 2 million user reviews for Facebook, WeChat, TikTok, Grammarly, COVID Symptom Study, and COVID-19 mobile apps. These reviews are extracted for the period between 1st of April 2020 and 31st of August 2020; this is the period where the COVID-19 apps started to be used across the globe.

| App Aspect | A user review containing... |
|---|---|
| **Privacy** | ... privacy or security related issues, e.g collecting or accessing users data, information, location, etc. |
| **Stability** | ... stability or failure related issues, e.g. crashes, freezes, bugs, etc. |
| **Advertising** | ... ads or commercials related issues, e.g frequent ad banners, pop-ups, etc. |
| **Requests** | ... user requests related issues, e.g. feature, bug fix, app updates requests, etc. |
| **Uninstallation** | ... users deleting or uninstalling the app because of privacy, stability issues, etc. |

TABLE II: App aspects used in our user reviews classification

Our tool is written in Python and uses GooglePlay and AppleStore open APIs to extract user reviews. Figure 3 illustrates the tool processing. If a review is not in English, it uses the Google Translate library to detect the language of all the reviews and translate them into English. It then classifies reviews into five different app aspects based on used keywords, chosen based on manually analysing over 23,000 user reviews.

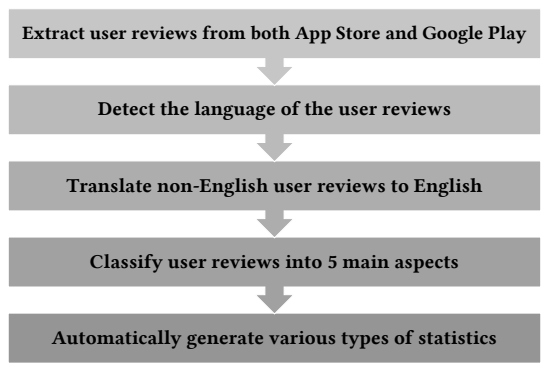

Fig. 3: Our automated tool technique

## IV. RESULTS

### A. RQ1 – How is personal data handled by COVID-19 and Social Media mobile apps?

We analysed the privacy policies and data use agreements for 42 official COVID-19 mobile apps. We did the same analysis for Facebook, TikTok, WeChat, and Grammarly to represent Social Media & Productivity applications.

*1) Is it an Open Source application?:* All Social Media and Productivity applications included in our study are not Open Source. There is no way to understand how the app actually works or what happens to the user's data unless you read the data use agreements. These are often unclear, not specific, and long, leading people to blindly accept the terms and conditions. As shown in Table III, 57% of COVID-19 applications included in our study are open source. Application code is publicly shared so it can be analysed to understand precisely how it works, including what algorithms and technologies are used, where data is saved, and what happens with the user's data.

*2) Are you forced to use it?:* All Social Media and Productivity mobile apps included in our study are voluntarily downloaded by the users. No one is forced to download them and their marketing is based on entertaining and providing services to the users. However, 16% of COVID-19 applications are mandatory to be downloaded by the users (citizens) as per their government's regulations. These regulations are set trying to stop or slow the spreading of this virus.

*3) Is your data used out of the app scope?:* As shown in Table III, only 30% COVID-19 apps included in our study can use app user data out of the scope of the application. However, this is nowhere near the extent of how Social Media & Productivity apps can use app user data. User data is used in different shapes and forms for Social Media & Productivity apps. They use the information they collect to promote their products or services, including to personalise features and content such as stories, ads, suggestions for the user. To make this personalisation unique and relevant for each user, they use user's behavioural attributes such as activities they are involved with, interests, connections, preferences based on the data they capture from and about the user and others including friends, relatives, etc. Moreover, they collect user behaviour and interaction with products or based on connections and chats with certain people, check in to different places, or even things they are connected to or shown interest in.

**Location-related information:** All Social Media & Productivity apps we included use location-related information, primarily the user's current location. Moreover, they collect user's current address and places they regularly visit, businesses they use, and people they are close by. Location-related information are usually based on precise device location (GPS). Sometimes users are tracked by their IP addresses, check-ins, or events information. In contrast, 82% of COVID-19 apps rely on Bluetooth contact tracing techniques and do not need to use GPS or IP address. So COVID-19 apps provide higher privacy for users since their location is never revealed or signed in. They record anonymous key codes and anonymised IDs are exchanged between nearby mobile devices. Hence, no personal data can be identified or transferred. Moreover, in a decentralised approach, the user data will only be stored and processed locally on user phones and will never be transferred to a central server. Bluetooth tracing is also more accurate than GPS for short distance tracing.

**Product research and development:** All Social Media & Productivity apps use user's information to develop, test, and improve their Products. User information is used in research and some users are not aware that their data is used in this way. However, COVID-19 apps do not use data for any product

| Name | Country | Open Source | Are you forced to use it? | Can your data be used out of the app scope? | Is your data deleted when it's no longer needed? | Does the app collect more data than it actually needed? | Is your identify masked? | Can the app access everything you type? | Can the app use your GPS? | Device signals |
|---|---|---|---|---|---|---|---|---|---|---|
| Facebook | All | N | N | Y | N | Y | N | N | Y | All |
| Grammarly | All | N | N | Y | N | Y | N | Y | Y | All |
| TikTok | All | N | N | Y | N | Y | N | N | Y | All |
| WeChat | All | N | N | Y | N | Y | N | N | Y | All |
| **Overall Stats** | Yes | 0% | 0% | 100.00% | 0% | 100.00% | 0% | 25.00% | 100.00% | 100.00% (ALL) |
|  | No | 100.00% | 100.00% | 0% | 100.00% | 0% | 100.00% | 75.00% | 0% | 0% |
| COVIDSafe | Australia | Y | N | N | Y | N | N | N | N | Bluetooth |
| Stopp Corona | Austria | Y | N | N | Y | N | Y | N | N | Bluetooth |
| BeAware | Bahrain | N | N | N | N | Y | N | N | Y | Bluetooth, Location |
| ViruSafe | Bulgaria | Y | N | N | Y | Y | Y | N | Y | Location |
| COVID Alert | Canada | Y | N | N | Y | N | Y | N | N | Bluetooth |
| Chinese health code system | China | N | Y | Y | N | Y | N | N | Y | Location |
| CovTracer | Cyprus | Y | N | Y | Y | N | Y | N | Y | Location |
| eRouska | Czech | Y | N | N | Y | N | Y | N | N | Bluetooth |
| Smittestopp | Denmark | N | N | N | Y | N | Y | N | N | Bluetooth |
| Seha Masr | Egypt | N | N | Y | N | Y | N | N | N | N/A |
| CareFiji | Fiji | Y | N | N | Y | N | Y | N | N | Bluetooth |
| Ketju | Finland | Y | N | N | Y | N | Y | N | N | Bluetooth |
| StopCovid | France | Y | N | N | Y | N | Y | N | N | Bluetooth |
| Corona-Warn-App | Germany | Y | N | N | Y | N | Y | N | N | Bluetooth |
| GH COVID-19 Tracker | Ghana | N | N | Y | N | Y | N | N | Y | Location |
| Beat COVID Gibraltar | Gibraltar | Y | N | N | Y | N | Y | N | N | Bluetooth |
| VirusRadar | Hungary | N | N | N | Y | N | Y | N | N | Bluetooth |
| Rakning C-19 | Iceland | Y | N | N | Y | N | Y | N | Y | Location |
| Aarogya Setu | India | Y | Y | Y | Y | Y | Y | N | Y | Bluetooth, Location |
| PeduliLindungi | Indonesia | N | N | Y | N | Y | N | N | Y | Bluetooth, Location |
| Covid Tracker | Ireland | Y | N | N | Y | N | Y | N | N | Bluetooth |
| HaMagen | Israel | Y | Y | N | Y | N | Y | N | Y | Location |
| Immuni | Italy | Y | N | N | Y | N | Y | N | N | Bluetooth |
| COCOA | Japan | N | N | N | Y | N | Y | N | N | Bluetooth |
| Shlonik | Kuwait | N | N | Y | N | Y | N | N | N | Location |
| MyTrace | Malaysia | Y | N | Y | N | Y | N | N | N | Bluetooth |
| CovidRadar | Mexico | N | N | Y | N | Y | N | N | N | Bluetooth |
| NZ COVID Tracer | New Zealand | Y | N | N | Y | Y | Y | N | N | Bluetooth |
| StopKorona | North Macedonia | N | N | N | N | N | Y | N | N | Bluetooth |
| Smittestopp | Norway | N | N | N | N | Y | N | N | Y | Bluetooth, Location |
| StaySafe | Philippines | N | N | Y | N | Y | N | N | N | Bluetooth |
| ProteGO | Poland | Y | N | N | Y | N | Y | N | N | Bluetooth |
| Ehteraz | Qatar | N | Y | Y | N | Y | N | N | Y | Bluetooth, Location |
| Tabaud | Saudi Arabia | N | N | Y | Y | Y | N | N | N | Bluetooth |
| Trace Together | Singapore | Y | N | N | Y | N | Y | N | N | Bluetooth |
| SwissCovid | Switzerland | Y | N | N | Y | N | Y | N | N | Bluetooth |
| MorChana | Thailand | Y | Y | N | N | Y | N | N | Y | Bluetooth, Location |
| E7mi | Tunisia | N | N | N | Y | Y | Y | N | N | Bluetooth |
| Hayat Eve Sığar | Turkey | N | Y | Y | N | N | N | N | Y | Bluetooth, Location |
| TraceCovid | UAE | N | Y | N | Y | N | N | N | N | Bluetooth |
| NHS COVID-19 | UK | Y | N | N | Y | N | Y | N | N | Bluetooth |
| BlueZone | Vietnam | Y | N | N | N | Y | Y | N | N | Bluetooth |
| **Overall Stats** | Yes | 57.14% | 16.67% | 30.95% | 66.05% | 42.86% | 42.86% | 0.00% | 33.33% | 82.93% (Bluetooth) |
|  | No | 42.86% | 83.33% | 69.05% | 33.33% | 57.14% | 57.14% | 100% | 66.67% | 17.07% |

TABLE III: Summary for Privacy and data use agreement between COVID-19, Social Media & Productivity apps. Some data has been extracted from [34]

research and development.

**Face recognition:** If the user turns on face recognition technology, this will give Social Media apps access user's photos, videos, and camera. No COVID-19 apps have that option.

**Ads and other sponsored content:** Social Media & Productivity apps use app user information, including their actions and interests to select ads and for ad personalisation. This also drives offers and other sponsored content that is shown to users. None of the COVID-19 apps has any ads.

*4) Is your data deleted when it's no longer needed?:* Any app should delete user data once it is no longer needed. Social Media & Productivity apps didn't delete user data until recently when they were forced to by new government regulations. Social Media shares user's information with people and accounts they share and communicate with. This often includes user material and data that others post or re-share about them, information about the active status of the user and their interaction with applications, websites and integration with third party apps and partners. For example, Facebook claims that it deletes a user's data within 90 days. However, they acknowledge that they will not be able to delete any data or information they have already shared with third-party apps, websites, games, and services. User data can also be shared with enforcement agencies and some have the power to force users to hand over their logs, whistleblower ids and sources. They can also identify and get the details for people attending a protest, leaking information to media and even release details of people in hiding from abusive partners, in witness protection, etc.

In contrast, more than 66% of the COVID-19 apps automatically delete all the app user data once it is no longer used. Some delete it immediately when a user deletes their account or no longer uses the app. Others delete it automatically after a specific period ranging between 1 to 4 weeks.

*5) Does the app collect more data than it actually needs?:* Social Media & Productivity apps are usually data-hungry and try to collect as much data as possible, whether or not the user knows about it. They track and monitor the online behaviour of users and their interacts with other users and apps. Social Media Productivity apps collect user device information, including device features and settings, signatures, signals, configuration data, network and connections, and cookie data. 42% of COVID-19 apps that we analysed collect more data that the app actually needs, but is much less than data mentioned above.

*6) Is your identity masked?:* None of the Social Media & Productivity apps mask users' data. 42% of COVID-19 apps mask their user's data to protect their privacy.

*7) Can the App access everything you type?:* Some Social Media & Productivity apps access everything you type, such as Grammarly. These apps will not work unless you allow them to record and monitor every single letter you type. This is a major ethical and security concern. On the other hand, none of the COVID-19 apps have this issue.

### B. RQ2 – What are the key issues raised by the users of COVID-19 and Social Media & Productivity apps as evidenced in their users' reviews?

We downloaded, analysed and classified 1,978,418 user reviews for Facebook, WeChat, TikTok, Grammarly, COVID Symptom Study, and 21 different COVID-19 mobile apps. All these user reviews were extracted from both GooglePlay and App Store. The number of analysed user reviews and the overall rating for each app is shown in Table IV.

| App | Average Rating | Number of reviews |
|---|---|---|
| **Facebook** | 3.49 | 817,980 |
| **TikTok** | 4.08 | 797,331 |
| **WeChat** | 2.63 | 84,309 |
| **Grammarly** | 3.85 | 28,027 |
| **COVID Symptom** | 4.68 | 28,306 |
| **All COVID apps** | 3.50 | 222,465 |

TABLE IV: Overall average rating and number of analysed user reviews

The extracted, nearly 2 million, user reviews have been classified into five main aspects. These aspects are privacy, stability, advertising, requests and uninstallation. The overall percentage of each analysed aspect among the total number of reviews are shown in Figure 4 and analysed below.

*1) Privacy:*
Despite the significant privacy issues involved within Social Media & Productivity apps, the percentage of users raising privacy issues within these apps were less than in COVID-19 apps as shown in Figure 4. However, in Figure 5, COVID-19 contact tracing apps user reviews discussing privacy-related issues were more positive compared to Social Media & Productivity apps, which was very surprising to us. This shows that **users who downloaded and used COVID-19 contact tracing apps were more satisfied from a privacy perspective** than users who were using Social Media & Productivity apps. Many users indicated in their reviews that their privacy in COVID-19 apps is much better handled than in Social Media apps. Moreover, some users mentioned that their data is not exposed and is more protected in COVID-19 apps compared to Social Media & Productivity apps. The users of the Grammarly productivity app were the most negatively concerned about the privacy issues within the app since the app tracks every single thing they type on their smartphones. The users of the COVID Symptom Study app were the most satisfied from a privacy perspective since the app does not track or trace them but gives them useful information about the virus. Example quotes:

> **Facebook User:** "Privacy don't exist!!! Please people read Facebook terms and conditions, no privacy , crazy !!"
> **Grammarly User:** "This app wants permission to collect my personal data such as passwords and credit card numbers. Take care for all of you who want to use this!"
> **COVIDSafe User:** "Amazing how many people allow Google, Apple, Facebook etc to access data yet will not trust an app that has the ability to assist in saving lifes."

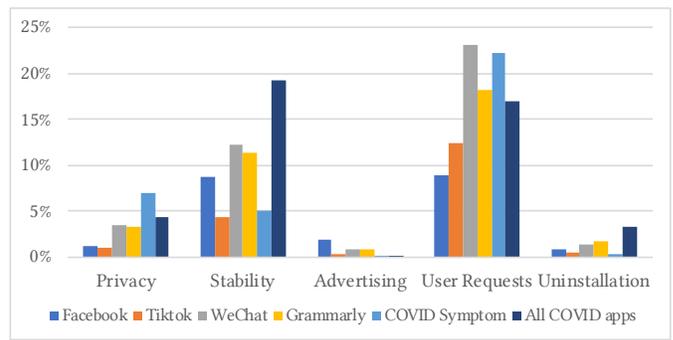

Fig. 4: Overall percentage for each analysed aspect across apps

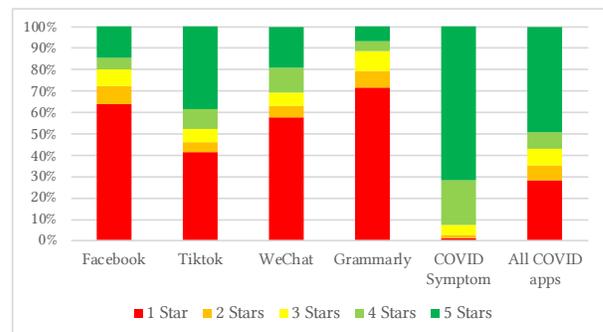

Fig. 5: Distribution of privacy issues across star ratings

*2) Stability:*
Stability issues reported in the user reviews within COVID-

19 contact tracing apps are significantly higher compared to Social Media & Productivity apps as reflected in Figure 4. The majority of these stability related issues were negative as shown in Figure 6. By manually reviewing the stability classified user reviews within COVID-19 apps, the majority of these reviews were reporting connectivity related issues with other Bluetooth devices, as most COVID-19 contact tracing apps are using Bluetooth for tracing, which affects the connectivity with other devices such as headphones and smartwatches. Moreover, many users reported that the app is running all the time in the background, which caused some freezing to their smartphones in addition to rapid battery drainage. On the other side, most of the stability issues reported in Social Media & Productivity apps were less severe such as bugs and long loading times. In Social Media & Productivity apps, the users reported in their reviews that by reinstalling or updating the app to the newer version, these issues are usually fixed. This actually shows that emerging COVID-19 apps usually have many severe bugs and instability issues, most likely because they have been developed, designed, and have limited testing in a short time span. In contrast, Social Media apps are way better tested and the developers have had a sufficient amount of time to fix major bugs and reported issues and to improve stability in their regular updates.

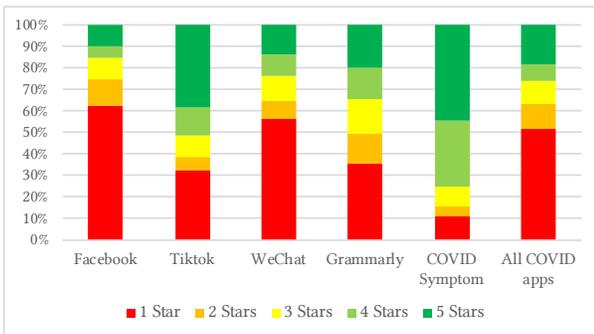

Fig. 6: Distribution of stability issues across star ratings

*3) Advertising*:
As shown in Figure 4, advertisement related issues were highly reported within Social Media & Productivity mobile apps compared to COVID-19 apps. None of the official COVID-19 apps run ads in their algorithms and that is why it been reflected in many COVID-19 app user reviews that they are happy with the COVID app is not showing ads like many other apps. For example, a user has written this review in Facebook app *"Ads, ads and more ads My timeline feels like it has more ads than posts. Stop with these advertisements. Very annoying and would like to just finish the videos or posts without seeing these"*. On the other side, a user wrote this review in Immuni app (Italian COVID-19 app) *"The only app on my phone without popup ads. 10/10"*. It is also shown in Figure 7, that user reviews containing advertising related issues in COVID apps were having better ratings compared to Social Media & Productivity mobile apps.

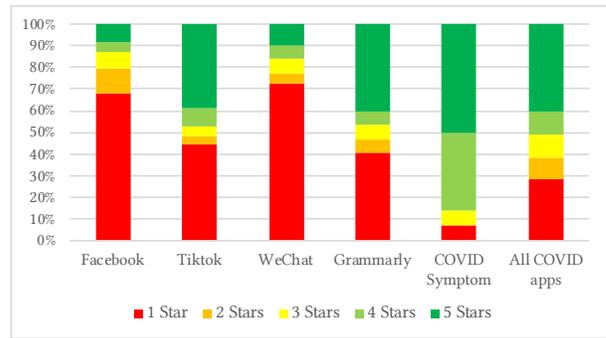

Fig. 7: Distribution of advertising issues across star ratings

*4) User Requests*:
Both COVID-19 and Social Media & Productivity mobile apps have lots of user requests, as presented in Figure 4. Around 20% of the user reviews of COVID-19 apps contained requests. By manually reviewing the users' requests in the classified reviews linked to COVID-19 apps, we determined that they were mainly asking for bug fixes including enhancing Bluetooth connectivity related issues and for enhancing accessibility within the apps. For example, a user has written this review in SmartStop app *"Oh dang dang, here I was expecting English version (at least) for such an important app. Not only dane, I've seen in Denmark I believe is difficult to use the app for the single language interface."*. It was surprising for us that most of these requests were very repetitive by different users, however, the issues raised and requested by users were not addressed or fixed in the next version. Most of COVID-19 apps didn't interact or give feedback for the user requests. This shows that the developers of most COVID-19 apps have miscommunication issues with their users. On the other hand, most of the user requests within Social Media & Productivity mobile apps were feature requests. The variations in star ratings across different apps are shown in Figure 8.

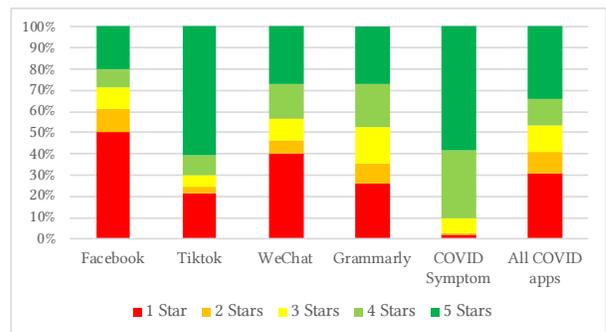

Fig. 8: Distribution of user requests issues across star ratings

*5) Uninstallation*:
COVID-19 contact tracing mobile apps had a higher rate of uninstallation rates than Social Media & Productivity apps. COVID-19 apps have high uninstallation rates as most applications are based on Bluetooth, which negatively affects the connectivity of the phone with other devices such as headphones, smartwatches, etc. Another reason for uninstalling

COVID-19 apps is due to some severe accessibility issues many users encounter. Some of COVID-19 apps are limited to be downloaded from a specific country store which makes these apps inaccessible for residents or visitors who are using a different country in their accounts. Moreover, another serious accessibility issue is that some of COVID-19 apps involve using a phone number for registration, however, these apps only support local numbers. Since users will not be able to register until confirming their local phone numbers, this issue makes these apps inaccessible since some of the users are using their international phone numbers, this lead at the end that users delete and uninstall the app. In addition, most COVID-19 apps only support one language, which is usually the local language for the country. This makes the app inaccessible for people who do not understand this language, this issue leads at the end that some users permanently delete or uninstall the app. On the other side, most of the uninstallation that occurs in Social Media & Productivity apps seems to be due to less severe issues such as bugs, long loading times, etc. It was commonly reported in the reviews that when users uninstall these applications or reinstall or update them to the newer version, these issues are usually fixed. The variations in star ratings across different apps are shown in Figure 9.

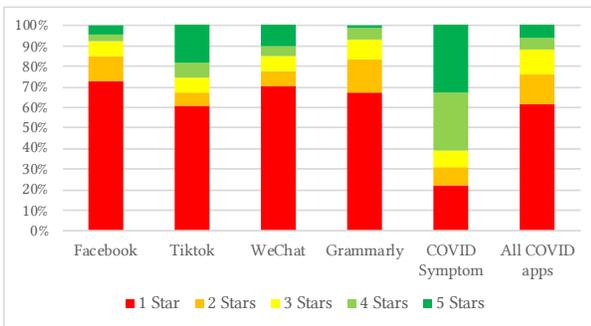

Fig. 9: Distribution of uninstallation issues across star ratings

### C. Threats to Validity

**Limited dataset of apps:** We have only included the official COVID-19 mobile apps in the study. In addition, we have represented Social Media & Productivity categories by Facebook, Tiktok, WeChat, and Grammarly apps.

**Limited information:** COVID-19 mobile apps have a significantly lower number of app reviews compared to Social Media & Productivity apps. Moreover, lots of users just submit a star rating without writing or leaving any feedback. Some users only write a short feedback and they do not profoundly express their user experience, issues, or concerns while using the app.

**Inaccurate translation:** Since we are using Google API for translating non-English user reviews to English. The translation can be inaccurate in minimal cases, which can lead to a wrong classification.

**Manual policy analysis:** One of the authors did a detailed analysis of app privacy and data usage policies, checked by another author. However these policies may not accurately reflect what the app actually does regarding user privacy and data usage. Closed source apps are very hard to check to see if they actually adhere to their claimed privacy and data usage policies.

**Automated review analysis:** We used a large phrase dataset to classify user reviews into the five categories and linked classified review with overall app rating. Our phrases may be inaccurate or miss some key words used in user reviews. We manually checked phrases used in over 23,000 reviews to build the dataset, reviews from many other types of apps, not just those of social media and COVID-19 apps.

## V. DISCUSSION

We compared the privacy policies and data use agreements of COVID-19, Social Media & Productivity apps. We analysed nearly 2 million user reviews for these apps and categorised them into five main aspects, privacy, stability, advertising, user requests, and uninstallation. Below we discuss our findings and implications for researchers and practitioners.

### A. Findings

**COVID-19 apps better support privacy and better limit data usage:** In our analysis of the privacy policies, terms & conditions, and data use agreements for COVID-19 and Social Media & Productivity apps included in our study, it was clearly shown that privacy and personal data are more violated and exposed in Social Media apps compared to most COVID-19 apps. Social Media & Productivity apps can use users' data and monitor their online behaviour and share these data and information with third parties. Users of Social Media apps can not really know or even control how their data can be stored, exchanged, transmitted, and with whom and when. Moreover, they do not have an easy way to fully delete their own data as their data is already shared with third parties. So in the end, the users of social media apps could not fully understand how the data is being used and for what and whether if their data is kept private or not. On the other hand, the data collected by most COVID-19 apps are usually encrypted, anonymous, and stored historically rather than in real-time just to protect the privacy of the users. The majority of COVID-19 apps do not record the real-location of the users' but just a record of anonymous key codes and anonymised IDs are exchanged between near mobile devices using Bluetooth. Hence, no personal data can be identified or even transferred between different users. Moreover, in the decentralised approach, the users' data will only be stored and processed locally on their phones and will never be transferred to a central server for processing, which makes the tracing procedure very secured.

**More COVID-19 app reviews talk about privacy issues:** Our analysis of app user reviews showed that privacy-related issues were raised by a higher rate in COVID-19 apps than in Social Media & Productivity apps included in our study. Despite the high rate, the majority of COVID-19 app users were actually satisfied with the privacy of the apps and were giving high ratings. On the other side, privacy-related issues were less raised in Social Media & Productivity apps but were mostly negative reviews when raised. In user reviews, many users made a comparison between COVID-19 and Social Media apps and mentioned that their data is not exposed and more protected in COVID-19 apps compared to Social Media & Productivity apps. In addition, users of COVID-19 apps were satisfied, as the COVID-19 apps do not have any advertisements. In contrast, the users of Social Media apps were very dissatisfied with the large number of ads included in these apps. Actually, all social media & productivity apps included in our study are tracking users' behaviour and online activity to make personalised ads for every single user, which does not occur in COVID-19 contact tracing apps.

**COVID-19 apps are much less stable and accessible:** Our user review analysis highlighted that most COVID-19 apps have many severe bugs, and instability issues compared to Social Media & Productivity apps. From a software engineering perspective, this is most likely happening as COVID-19 apps are developed, designed, and tested in a short time span, while on the other side, Social Media apps are way better tested and the developers have plenty of time to fix any bugs and issues and improve stability in their regular updates. The majority of uninstallation happening in COVID-19 apps was because of stability or accessibility serious problems. Some of COVID-19 apps are inaccessible due to several reasons. These accessibility issues prevented users from using these apps. Most of these accessibility issues were very repetitive in the reviews, however, they were not fixed in the updates of the apps. This shows that the developers of most COVID-19 apps have miscommunication issues with their users.

### B. Implications for Researchers and Practitioners

Many people have lots of privacy misconceptions about COVID-19 apps and that is probably why they have low download and adoption rates. Our user reviews analysis showed that the majority of people who decided to use COVID-19 apps are actually satisfied with its privacy, but more dissatisfied with its stability, accessibility, and bugs. To fight this pandemic more effectively, health officials and technologists need to raise awareness for the public about the mechanism of how contact tracing apps actually work in a simple way so people with weak technology background can understand. This will allow the public to get their information from health officials and technologists not from media and will encourage them to download and start using these apps.

From a software engineering perspective, more research needs to be done on how to design and implement COVID-19 apps – and indeed any future public service apps that need to be rolled out quickly – in a way that allows users to be sure that their privacy is not violated. This will help health officials to collect data from users and at the same time, users become satisfied as their privacy is not violated. For example, apps should not ask for location, contacts, photos, camera, etc access as this worries users. Most COVID-19 apps need major accessibility and stability improvements.

Users need to be more aware that their privacy will be severely violated if they decide to use many common Social media apps and there is no way to prevent this as discussed in the paper. Further techniques to better preserve user privacy and limit data collection and use are needed for third party apps, from phone platforms and app analysis. App developers and organisations need to reconsider their lack of support of user privacy and data collection and mis-use. Ultimately, further government regulation about ensuring user privacy and limiting collected data mis-used may even be required.

## VI. RELATED WORK

A number of studies have investigated and analysed user reviews for mobile apps. In [35], an exploratory study was performed on what could impact the users to write an app review. In [36], they investigated different factors that affect user reviews, such as the correlation between the rating, price, and the number of downloads. The authors of [37], proposed a computational structure by adopting a semi-supervised algorithm in order to extract and rank insightful reviews. In [38], the authors developed a semi-automated keyword-based approach for mining user opinions in their reviews.

Multiple studies have reviewed privacy policies, terms & conditions, and data use agreements for some social media platforms [39]–[41]. Some work was done in investigating COVID-19 apps. In [42], they investigated the ad transparency mechanisms in social media. In [34], [43], they investigated ethical and privacy aspects for COVID-19 apps.

So far, no work is done on mining and analysing both user reviews and privacy policies, terms and conditions, and data use agreements at the same time. In our work, we investigated privacy policies of COVID-19 apps vs social media and productivity apps and showed COVID-19 apps have much more stringent privacy and data usage policies, and some are open source for inspection. User review analysis indicated many users appreciated this despite media claims to the contrary. However, COVID-19 apps have many stability and resource usage problems that must be better addressed.

We plan to conduct surveys and actual interviews with people who refuse to use COVID-19 contact tracing mobile apps and at the same time use Social Media apps. These surveys will help us better understand why so many people are voluntarily using Social Media apps without seeming to have concerns but are very concerned regarding using COVID-19 apps. We wonder if people trust private companies more than governments to handle their personal data.

## VII. SUMMARY

We analysed the privacy policies, terms & conditions and data use agreements for the most commonly used COVID-19, Social Media & Productivity apps. Moreover, we developed a tool to extract and analyse nearly 2 million user reviews for these apps. From our analysis, we found that, relative to COVID-19 apps, Social Media & Productivity apps have far more privacy and ethical problems. Many users indicated in their user reviews that their privacy in COVID-19 apps is much better treated than in social media apps. On the other hand, relative to most Social Media apps, most of the

COVID-19 apps are much less stable and accessible. Our study findings indicate that health officials and technologists need to better raise awareness among individuals about the behaviour and trustworthiness of COVID-19 apps in order to effectively tackle this pandemic. This will allow individuals to better understand and encourage them to download and use these apps. In addition, COVID-19 apps need major accessibility and stability enhancements to enable a broader variety of users from various societies and cultures to access these apps.

ACKNOWLEDGMENT

This work is partially supported by Australian Research Council Laureate Fellowship FL190100035 and Monash FIT.

REFERENCES

[1] T. Nabity-Grover, C. M. Cheung, and J. B. Thatcher, "Inside out and outside in: How the covid-19 pandemic affects self-disclosure on social media," *Int. Journal. Information Management*.
[2] J. Alanko and E. Laaksonen, "Social media marketing and growth study for an esports and gaming business: Case company x." 2020.
[3] L. L. T. Statista. (2020) Number of monthly active wechat users from 2nd quarter 2011 to 2nd quarter 2020. [Online]. Available: https://www.statista.com/statistics/255778/number-of-active-wechat-messenger-accounts
[4] R. Sunhare and Y. Shaikh, "Study of security vulnerabilities in social networking websites," *Int. Journal. Management, IT and Engineering*, vol. 9, no. 6, pp. 278–291, 2019.
[5] F. Kurniasari and J. N. Prihanto, "Implementation of productivity apps to increase financial inclusion in peer-to-peer lending platform," in *29th EBES CONFERENCE–LISBON PROCEEDINGS VOLUME*, p. 217.
[6] Grammarly.com. (2020) Grammarly for chrome. [Online]. Available: https://chrome.google.com/webstore/detail/grammarly-for-chrome/
[7] Y. K. Dwivedi, D. L. Hughes, C. Coombs, I. Constantiou, Y. Duan, J. S. Edwards, B. Gupta, B. Lal, S. Misra, P. Prashant *et al.*, "Impact of covid-19 pandemic on information management research and practice: Transforming education, work and life," *Int. Journal. Information Management*, p. 102211, 2020.
[8] D. Pal, V. Vanijja, and S. Patra, "Online learning during covid-19: Students' perception of multimedia quality," in *Proc. 11th Int. Conf. on Advances in Information Technology*, 2020, pp. 1–6.
[9] H. Cho, D. Ippolito, and Y. W. Yu, "Contact tracing mobile apps for covid-19: Privacy considerations and related trade-offs," *arXiv preprint arXiv:2003.11511*, 2020.
[10] S. A. Lauer, K. H. Grantz, Q. Bi, F. K. Jones, Q. Zheng, H. R. Meredith, A. S. Azman, N. G. Reich, and J. Lessler, "The incubation period of coronavirus disease 2019 (covid-19) from publicly reported confirmed cases: estimation and application," *Annals of internal medicine*, vol. 172, no. 9, pp. 577–582, 2020.
[11] J. Watkins, "Preventing a covid-19 pandemic," 2020.
[12] P. H. O'Neill. (2020) No, coronavirus apps don't need 60% adoption to be effective. [Online]. Available: https://www.technologyreview.com/2020/06/05/1002775/covid-apps-effective-at-less-than-60-percent-download
[13] O. University. (2020) Digital contact tracing can slow or even stop coronavirus transmission and ease us out of lockdown. [Online]. Available: https://www.research.ox.ac.uk/Article/2020-04-16-digital-contact-tracing-can-slow-or-even-stop-coronavirus-transmission-and-ease-us-out-of-lockdown
[14] E. Mbunge, "Integrating emerging technologies into covid-19 contact tracing: Opportunities, challenges and pitfalls," *Diabetes & Metabolic Syndrome: Clinical Research & Reviews*, 2020.
[15] J. Abeler, M. Bäcker, U. Buermeyer, and H. Zillessen, "Covid-19 contact tracing and data protection can go together," *JMIR mHealth and uHealth*, vol. 8, no. 4, p. e19359, 2020.
[16] P. International. (2020) Apps and covid-19. [Online]. Available: https://privacyinternational.org/examples/apps-and-covid-19/
[17] H.-T. Chen, "Revisiting the privacy paradox on social media with an extended privacy calculus model: The effect of privacy concerns, privacy self-efficacy, and social capital on privacy management," *American behavioral scientist*, vol. 62, no. 10, pp. 1392–1412, 2018.
[18] Worldometers. (2020) Coronavirus live update. [Online]. Available: https://www.worldometers.info/coronavirus/
[19] T. T. Le, Z. Andreadakis, A. Kumar, R. G. Roman, S. Tollefsen, M. Saville, and S. Mayhew, "The covid-19 vaccine development landscape," *Nat Rev Drug Discov*, vol. 19, no. 5, pp. 305–306, 2020.
[20] J. Morley, J. Cowls, M. Taddeo, and L. Floridi, "Ethical guidelines for covid-19 tracing apps," 2020.
[21] Facebook. (2020) Data policy. [Online]. Available: https://www.facebook.com/policy.php
[22] TikTok. (2020) Data policy. [Online]. Available: https://www.tiktok.com/legal/privacy-policy
[23] WeChat. (2020) Data policy. [Online]. Available: https://www.wechat.com/en/privacy_policy.html
[24] Grammarly. (2020) Data policy. [Online]. Available: https://www.grammarly.com/privacy-policy
[25] Amnesty. (2020) Bahrain, kuwait and norway contact tracing apps among most dangerous for privacy. [Online]. Available: https://www.amnesty.org/en/latest/news/2020/06/bahrain-kuwait-norway-contact-tracing-apps-danger-for-privacy/
[26] TheNewYorkTimes. (2020) Major security flaws found in south korea quarantine app. [Online]. Available: https://www.nytimes.com/2020/07/21/technology/korea-coronavirus-app-security.html
[27] A. Segal, "When china rules the web: Technology in service of the state," *Foreign Aff.*, vol. 97, p. 10, 2018.
[28] Google. (2020) Open letter to australians. [Online]. Available: https://about.google/intl/ALL_au/google-in-australia/aug-17-letter/
[29] T. S. M. Herald and T. Age. (2020) Tiktok, wechat to face australian social media security investigation. [Online]. Available: https://www.smh.com.au/politics/federal/tiktok-wechat-to-face-australian-social-media-security-investigation-20200717-p55d3y.html
[30] M. A. Beam, M. J. Hutchens, and J. D. Hmielowski, "Facebook news and (de) polarization: reinforcing spirals in the 2016 us election," *Information, Communication & Society*, vol. 21, no. 7.
[31] Theguardian.com. (2020) Trump to ban us downloads of tiktok and wechat. [Online]. Available: https://www.theguardian.com/technology/2020/sep/18/tiktok-ban-us-downloads-wechat-latest
[32] COVIDSafe. (2020) Data policy. [Online]. Available: https://covidsafe.gov.au/privacy-policy.html
[33] Corona-Warn-App. (2020) Data policy. [Online]. Available: https://www.coronawarn.app/en/faq/
[34] P. O'Neill, T. Ryan-Mosley, and B. Johnson, "A flood of coronavirus apps are tracking us. now it's time to keep track of them," 2020.
[35] R. Vasa, L. Hoon, K. Mouzakis, and A. Noguchi, "A preliminary analysis of mobile app user reviews," in *Proc. 24th Australian Computer-Human Interaction Conference*, 2012, pp. 241–244.
[36] C. Iacob, V. Veerappa, and R. Harrison, "What are you complaining about?: a study of online reviews of mobile applications," in *27th International BCS Human Computer Interaction Conference (HCI 2013) 27*, 2013, pp. 1–6.
[37] N. Chen, J. Lin, S. C. Hoi, X. Xiao, and B. Zhang, "Ar-miner: mining informative reviews for developers from mobile app marketplace," in *Proc. 36th Int. Conf. on software engineering*, 2014, pp. 767–778.
[38] P. M. Vu, T. T. Nguyen, H. V. Pham, and T. T. Nguyen, "Mining user opinions in mobile app reviews: A keyword-based approach (t)," in *2015 30th IEEE/ACM Int. Conf. on Automated Software Engineering (ASE)*. IEEE, 2015, pp. 749–759.
[39] J. M. Such and M. Rovatsos, "Privacy policy negotiation in social media," *ACM Transactions on Autonomous and Adaptive Systems (TAAS)*, vol. 11, no. 1, pp. 1–29, 2016.
[40] S. Zimmeck, Z. Wang, L. Zou, R. Iyengar, B. Liu, F. Schaub, S. Wilson, N. M. Sadeh, S. M. Bellovin, and J. R. Reidenberg, "Automated analysis of privacy requirements for mobile apps." in *NDSS*, 2017.
[41] L. Parker, V. Halter, T. Karliychuk, and Q. Grundy, "How private is your mental health app data? an empirical study of mental health app privacy policies and practices," *Int. Journal. Law and Psychiatry*, vol. 64, pp. 198–204, 2019.
[42] A. Andreou, G. Venkatadri, O. Goga, K. Gummadi, P. Loiseau, and A. Mislove, "Investigating ad transparency mechanisms in social media: A case study of facebook's explanations," 2018.
[43] R. A. Fahey and A. Hino, "Covid-19, digital privacy, and the social limits on data-focused public health responses," *Int. Journal. Information Management*, p. 102181, 2020.